\documentclass[sigconf]{acmart}
\newcommand{\isACM}{true}
\usepackage{subcaption}
\usepackage{tabularx}
\usepackage{booktabs}

\AtBeginDocument{%
  }

\makeatletter
\let\@acmbadge\@empty               
\makeatother

\acmConference[]{}{}{}
\acmDOI{}
\acmISBN{}
\acmYear{}
\acmPrice{}

\AtBeginDocument{\pagestyle{plain}}

\author{Syafiq Al Atiiq}
\email{syafiq_al.atiiq@eit.lth.se}
\affiliation{%
  \institution{Lund University}
  \city{Lund}
  \country{Sweden}
}

\author{Chun Zhou}
\email{ch7871zh-s@student.lu.se}
\affiliation{%
  \institution{Lund University}
  \city{Lund}
  \country{Sweden}
}

\author{Christian Gehrmann}
\email{christian.gehrmann@eit.lth.se}
\affiliation{%
  \institution{Lund University}
  \city{Lund}
  \country{Sweden}
}

\begin{document}

 \title{Dissecting the Black Box: Circuit-Level Analysis of LLM Vulnerability Detection}



\begin{abstract}
Large language models (LLMs) can detect software vulnerabilities, but how do they actually identify vulnerable code? We address this question using mechanistic interpretability; analyzing the internal computations of a neural network to understand its reasoning process.

Using Circuit Tracer on Gemma-2-2b, we trace the computational pathways activated when the model classifies 472 C/C++ code samples as vulnerable or safe. Our analysis reveals a surprising finding: the model primarily relies on \textit{safety detectors}, attention heads that recognize safe coding patterns, rather than directly detecting vulnerability signatures. When these safety detectors fail to activate, the model classifies code as vulnerable.

We identify the critical neural components: specific attention heads in early layers (L5, L7) that focus on safety patterns, and Multilayer Perceptron (MLP) neurons in Layer 7 that encode vulnerability-related features. Ablation experiments confirm their causal role; removing Layer 11 drops vulnerability detection accuracy from 100\% to 6\%, while ablating just 20 neurons in Layer 7 reduces it by 50\%.

Our findings show that LLM vulnerability detection uses sparse, interpretable circuits (only 16\% of model capacity), enabling circuit-level explanations for security predictions and targeted improvements to detection systems.
\end{abstract}

\ifdefined\isACM
\fi


\ifdefined\isACM
\maketitle
\fi

\section{Introduction}

Large Language Models (LLMs) have demonstrated remarkable capabilities in software vulnerability detection, often matching or exceeding human expert performance~\cite{steenhoek2023codellm,chakraborty2022deep}. Yet despite this progress, most research focuses exclusively on improving detection accuracy rather than understanding \textit{how} these models identify vulnerable code. This limits our ability to debug failures, improve robustness, and build trust in security-critical applications. When an LLM flags code as vulnerable, practitioners cannot verify whether the model identified a genuine security flaw or exploited spurious correlations in the training data.

We address this gap using mechanistic interpretability~\cite{elhage2021mathematical,olah2018building}, which analyzes the internal computations of neural networks to understand their reasoning. Tools like Circuit Tracer~\cite{circuit_tracer} enable researchers to trace computational pathways through transformer layers, identifying which specific attention heads, neurons, and circuits are responsible for particular behaviors. This level of analysis goes beyond traditional black-box evaluation to reveal the actual computational mechanisms underlying model predictions.

In this work, we conduct the first systematic study of the computational circuits responsible for vulnerability detection in LLMs. Using Circuit Tracer, we analyze how vulnerable and non-vulnerable code samples flow through transformer layers, identifying the specific components that enable vulnerability recognition. Our investigation reveals that vulnerability detection is not a distributed, holistic process but rather relies on discrete neural circuits concentrated in specific layers.

Our key contributions include:

\begin{itemize}
\item \textbf{Safety detector mechanism}: We discover that LLMs detect vulnerabilities indirectly through ``safety detectors,'' attention heads that recognize safe coding patterns. Vulnerability classification occurs when these detectors fail to activate, a fundamentally different mechanism than direct vulnerability recognition.
\item \textbf{Critical component identification}: We pinpoint the specific neural components responsible for detection: attention heads L5-H2 and L7-H6 for safety pattern recognition, and 20 specialized neurons in Layer 7 that encode vulnerability-related features.
\item \textbf{Causal validation}: Through ablation experiments, we demonstrate that these components are causally necessary. Removing Layer 11 drops vulnerability detection from 100\% to 6\%, while ablating just 20 neurons in Layer 7 reduces it by 50\%.
\item \textbf{Sparse, modular circuits}: We show that vulnerability detection uses only 16.1\% of model capacity, with a hierarchical structure that mirrors human code analysis: syntactic processing in early layers converging to abstract decisions in late layers.
\end{itemize}

Our code is available at this repository\footnote{\url{https://anonymous.4open.science/r/LLMvul-02E6/}}.

\section{Related Work}

\subsection{LLM-based Vulnerability Detection}

Recent years have witnessed significant progress in applying large language models to software security analysis. CodeBERT~\cite{feng2020codebert} and similar pre-trained models~\cite{niu2022spt,svyatkovskiy2020intellicode} have shown promising results in detecting various types of vulnerabilities including buffer overflows, injection attacks, and cryptographic misuse. Empirical studies~\cite{steenhoek2023codellm,chakraborty2022deep} have demonstrated that deep learning models can achieve competitive performance with traditional static analysis tools, while neural approaches like VulDeePecker~\cite{li2018vuldeepecker} and Devign~\cite{zhou2019devign} have established benchmarks for vulnerability detection.

DiverseVul~\cite{diversevul} represented a significant step forward by curating a large-scale dataset of 18,945 vulnerable functions spanning 150 CWEs and 330,492 non-vulnerable functions from 7,514 commits across 295 projects. Their comprehensive study of 11 model architectures demonstrated that code language models outperform Graph Neural Networks as training data increases, though even the best models achieved only 11.94\% F1 score, highlighting persistent generalization challenges.

However, subsequent work revealed significant limitations in existing benchmarks including DiverseVul. PrimeVul~\cite{ding2025primevul} conducted a comprehensive analysis showing that prior datasets suffer from noisy labels (24-60\% accuracy), high duplication rates, and evaluation methodologies that substantially overestimate model performance. For instance, StarCoder2~\cite{lozhkov2024starcoder2stackv2} achieved 68.26\% F1 on BigVul~\cite{bigvul} but only 3.09\% on the more rigorous PrimeVul benchmark, exposing a large gap between reported performance and real-world effectiveness. These findings show the need for both higher-quality benchmarks and deeper understanding of model internals.

Most existing work focuses on improving detection accuracy rather than understanding the internal mechanisms of these models. This limits our ability to debug failures, improve robustness, and build trust in security-critical applications. Our work addresses this gap by providing the first mechanistic analysis of how LLMs process vulnerability-related code patterns.

\subsection{Mechanistic Interpretability}

Mechanistic interpretability seeks to understand neural networks by analyzing their internal representations and computation patterns~\cite{olah2018building}. Recent breakthroughs in transformer interpretability~\cite{elhage2021mathematical} have revealed that these models learn decomposable circuits that perform specific computational tasks. Researchers have identified specialized mechanisms such as induction heads~\cite{olsson2022context} and interpretable features through sparse autoencoders~\cite{cunningham2023sparse,templeton2024scaling}.

In the context of transformer models, researchers have identified specialized circuits for tasks such as indirect object identification~\cite{wang2022interpretability} and various linguistic phenomena. These studies typically involve analyzing attention patterns, probing internal representations, and examining how information flows through different layers of the transformer architecture~\cite{vaswani2017attention,raghu2021vision}.

\subsection{Circuit Discovery Across Domains}

Circuit-level analysis has proven valuable across diverse cognitive tasks beyond language modeling. Olsson et al.~\cite{olsson2022induction} identified induction heads (two-layer attention circuits that implement in-context learning through pattern completion). Wang et al.~\cite{wang2022ioi} reverse-engineered the indirect object identification (IOI) circuit in GPT-2, discovering 26 attention heads organized into seven functional classes. In the domain of factual knowledge, Meng et al.~\cite{meng2022rome} demonstrated that facts are stored in mid-layer Multilayer Perceptron (MLP) modules and can be edited via rank-one updates (ROME). Nanda et al.~\cite{nanda2023grokking} provided mechanistic explanations for the ``grokking'' phenomenon, showing how transformers learn discrete Fourier transforms for modular arithmetic. For safety-relevant behaviors, Zou et al.~\cite{zou2023repeng} introduced representation engineering to monitor and manipulate high-level cognitive phenomena such as honesty and harmlessness, while Marks and Tegmark~\cite{marks2023truth} demonstrated that LLMs linearly represent truth and falsehood in their activation space. More recently, Marks et al.~\cite{marks2024sparse} developed sparse feature circuits using interpretable causal graphs, and Bereska and Gavves~\cite{bereska2024mech} provided a comprehensive review connecting mechanistic interpretability to AI safety. Our work extends this line of research to the security domain, applying circuit discovery techniques to understand how LLMs detect software vulnerabilities.

\subsection{Activation Analysis in Deep Learning}

Activation magnitude and sparsity patterns have been used to understand neural network behavior across various domains~\cite{li2016visualizing,ghorbani2019towards}. The L0 norm, measuring the number of non-zero activations, provides insights into which neurons are engaged for specific tasks. Previous work has shown that different tasks can produce distinct activation signatures, making sparsity analysis a powerful tool for interpretability research. Recent work in representation engineering~\cite{zou2023representation} has further demonstrated how activation patterns can be analyzed and modified to understand model behavior.

Our work builds upon these foundations by applying activation analysis specifically to the domain of vulnerability detection, revealing novel insights into how LLMs process security-relevant code patterns.

\section{Methodology}

\subsection{Experimental Setup}

\textbf{Model Architecture.} We utilize Gemma-2-2b, a 2-billion parameter language model from Google's Gemma family, as our base model for vulnerability detection. The model employs a transformer architecture with 26 layers, 8 attention heads per layer, and a hidden dimension of 2,304. We access the model through the Circuit Tracer framework via the ReplacementModel interface, which enables fine-grained attribution analysis of model activations and feature-level interpretability. We use the pre-trained model without additional fine-tuning to analyze its inherent vulnerability detection capabilities.

\textbf{Hardware/Software Environment.} All experiments are conducted on an HPC cluster using a single NVIDIA A100 (80GB) GPU. The software stack includes Circuit Tracer framework (v0.1.0) with Gemma transcoder support, PyTorch 2.0, and TransformerLens for activation extraction.

\textbf{Dataset.} We use a balanced subset of $N=472$ samples from PrimeVul~\cite{ding2025primevul}, a high-quality vulnerability detection benchmark with 86-92\% label accuracy. We select 236 vulnerable and 236 non-vulnerable C/C++ functions in a strict 1:1 ratio to eliminate class imbalance bias that could confound circuit analysis. The subset spans 9 major CWE categories including buffer errors (CWE-119, 125, 787), input validation (CWE-20), null pointer dereferences (CWE-476), use-after-free (CWE-416), and integer overflows (CWE-190). This size enables intensive per-sample circuit tracing while maintaining diversity across vulnerability types; a key requirement for mechanistic interpretability studies where computational cost scales quadratically with dataset size.

\subsection{Circuit Tracer Instrumentation}

We employ Circuit Tracer~\cite{circuit_tracer} (version 0.1.0) for mechanistic interpretability analysis. Circuit Tracer is a tool designed to trace computational pathways through neural networks, enabling fine-grained analysis of how information flows through transformer layers.

Circuit Tracer uses gradient-based attribution to identify which model components contribute most significantly to the output. We configure the tracer with an attribution threshold of 0.01 to filter noise while preserving meaningful signal paths. For each sample, we perform full circuit tracing across all 26 layers, capturing attention patterns, MLP activations, and residual stream contributions.

For each code sample, Circuit Tracer allows us to:
\begin{enumerate}
\item \textbf{Trace activation paths}: Identify which neurons and attention heads contribute most significantly to the vulnerability prediction
\item \textbf{Compute attribution scores}: Quantify the importance of specific components in the decision-making process
\item \textbf{Extract circuit structures}: Map the computational graph of features that lead to vulnerability detection
\item \textbf{Perform causal interventions}: Test whether identified components are necessary for correct predictions
\end{enumerate}

\subsection{L0 Norm Analysis}

To quantify the computational density and feature engagement during vulnerability detection, we employ L0 norm analysis on the model's latent representations. Following the methodology established in recent mechanistic interpretability research ~\cite{pach2025sparse,lieberum2024gemma},the L0 norm provides insights into which neurons are actively participating in the computation for a given input. Formally, for a hidden state vector $\mathbf{h} \in \mathbb{R}^d$ at layer $\ell$, we compute:

\begin{equation}
\|\mathbf{h}_\ell\|_0 = |\{i : h_{\ell,i} \neq 0\}|
\end{equation}

where $h_{\ell,i}$ represents the activation of neuron $i$ at layer $\ell$. This metric allows us to quantify the sparsity patterns across different layers and compare how vulnerable versus non-vulnerable code samples activate the model's internal representations.

In our implementation, we utilize the Circuit Tracer framework  to capture these activations from the MLP output projections. To ensure numerical stability and filter out background noise, we apply a minimal threshold of $|h_i| > 10^{-6}$ for an activation to be considered "active". By comparing the L0 trends across different layers, we can distinguish between sparse, pattern-matching computations in early layers and more dense, distributive reasoning processes in the intermediate layers where vulnerability signatures are primarily processed. This metric allows us to characterize the "computational footprint" of specific security flaws compared to safe coding patterns.

\subsection{Circuit-Level Analysis}

\subsubsection{Attention Head Analysis}
To dissect the model's focus mechanism, we analyze the attention patterns of correctly classified samples (TP and TN). We propose a composite Importance Score ($I_h$) to quantify the functional divergence of each head $h$ between vulnerable and safe contexts:
\begin{equation}
I_h = (\bar{m}{TP} - \bar{m}{TN}) + \lambda (\bar{H}{TN} - \bar{H}{TP})
\end{equation}
where $\bar{m}$ represents the mean maximum attention weight and $\bar{H}$ denotes the attention entropy. This metric rewards heads that exhibit sharper focus (lower entropy) and higher intensity in one class over the other. To mitigate the ``attention sink'' phenomenon common in causal LLMs, we masked the beginning-of-sequence (BOS) token during calculation. A negative $I_h$ indicates the head acts as a Safety Detector, activating strongly only in the presence of safe code patterns. We use $\lambda = 0.5$ as a scaling factor to appropriately balance the contributions of attention entropy and intensity.

\subsubsection{MLP Neuron Analysis}
While attention heads facilitate information routing, MLP layers are hypothesized to serve as key-value memories that encode specific semantic features. We developed a probing framework to identify neurons specialized for vulnerability detection in intermediate layers (specifically layers 6, 7, 10, and 11).

To quantify the specialization of each neuron $n$, we propose a \textit{Vulnerability Selectivity Score} ($S_n$). Let $a_n(x)$ denote the post-activation value of neuron $n$ for a given input sample $x$. We compute the divergence in activation between the set of vulnerable samples ($X_{VUL}$) and safe samples ($X_{SAFE}$):
\begin{equation}
    S_n = \frac{1}{|X_{VUL}|} \sum_{x \in X_{VUL}} a_n(x) - \frac{1}{|X_{SAFE}|} \sum_{x \in X_{SAFE}} a_n(x)
\end{equation}
Neurons with a high positive $S_n$ are identified as \textbf{Vulnerability-Selective}, indicating they systematically fire in the presence of vulnerability patterns while remaining suppressed in safe contexts.

We rank all neurons in the target layers by $S_n$ and extract the top-$k$ (where $k=20$) candidates. To verify their role, we construct a contrastive activation matrix $\mathbf{M} \in \mathbb{R}^{k \times (N_{VUL} + N_{SAFE})}$. We visualize this matrix as a heatmap to inspect for a ``detectable boundary,'' namely a distinct transition from high activation (vulnerable samples) to inhibition (safe samples). This visual confirmation ensures that the identified neurons function as stable feature detectors rather than responding to transient noise.

\subsubsection{Causal Intervention Experiments}
To establish causal relationships between identified components and vulnerability detection, we design three intervention experiments: (1) \textit{Layer-wise Ablation} replaces activations at target layers (6, 7, 10, 11) with their mean values computed across the dataset, effectively removing layer-specific information while preserving computational flow. (2) \textit{Neuron-level Ablation} zeros out activations of the top-$k$ vulnerability-selective neurons identified by our selectivity metric, measuring the impact on detection accuracy. (3) \textit{Attention Head Knockout} disables specific attention heads by setting their output to zero, testing whether the identified safety detectors are necessary for correct classification.

\subsection{Data Collection}

We instrument the transformer model using Circuit Tracer to capture comprehensive activation data at each layer during forward passes. For each input sample, we record:

\begin{itemize}
\item Layer-wise L0 norm values across all 26 transformer layers
\item Attention head activation patterns (per-head, per-layer), including attention weights and entropy
\item MLP neuron activation magnitudes at intermediate layers (6, 7, 10, 11)
\item Attribution scores from Circuit Tracer indicating component importance
\item Input sample metadata (vulnerability type, CWE classification)
\item Final prediction confidence scores and logit distributions
\end{itemize}

This comprehensive data collection enables us to analyze both aggregate patterns across the dataset and individual sample trajectories through the model's computational graph.

\section{Results}

\subsection{Activation Pattern Analysis}

\begin{figure}
  \includegraphics[width=\columnwidth]{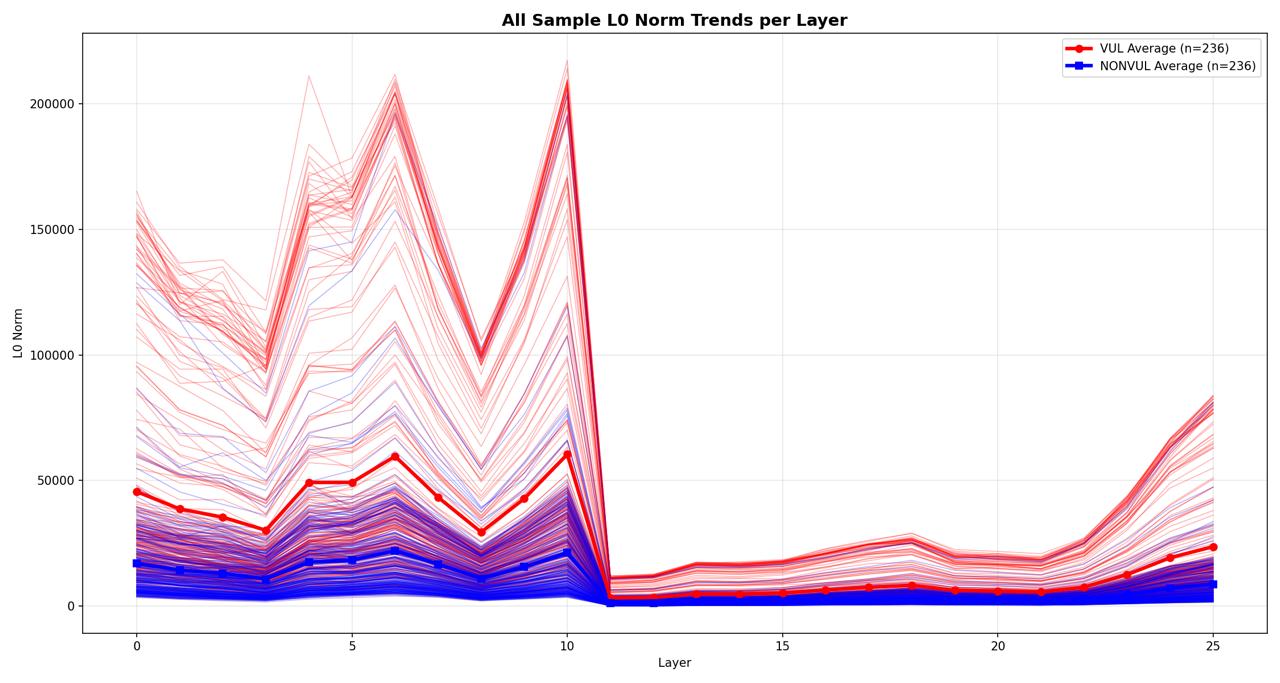}
  \caption{L0 norm trends across transformer layers for vulnerable (red) and non-vulnerable (blue) code samples. Each line represents an individual sample, with bold lines showing the average for each category (n=236 each).}
  \label{fig:l0_trends}
  \ifdefined\isACM
  \Description{A line plot showing L0 norm values on the y-axis (ranging from 0 to over 200,000) and layer numbers on the x-axis (0 to 25). Red lines represent vulnerable samples and blue lines represent non-vulnerable samples. The plot shows vulnerable samples have consistently higher L0 norms with characteristic spikes around layers 6-7 and 10-11.}
  \fi
\end{figure}

Figure~\ref{fig:l0_trends} presents our key finding: vulnerable and non-vulnerable code samples exhibit markedly different activation patterns across transformer layers. The analysis reveals several important observations:

\begin{enumerate}
\item \textbf{Magnitude Differences}: Vulnerable samples consistently produce higher L0 norm values across all layers, with average values ranging from approximately 25,000-220,000 compared to 5,000-15,000 for non-vulnerable samples.

\item \textbf{Layer-specific Spikes}: Both vulnerable and non-vulnerable samples show characteristic peaks in layers 6-7 and 10-11, suggesting these layers play crucial roles in feature extraction and vulnerability pattern recognition.

\item \textbf{Variance Patterns}: Vulnerable samples exhibit significantly higher variance in activation patterns, indicating more diverse computational pathways are engaged when processing security-relevant code.

\item \textbf{Late Layer Behavior}: In the final layers (20-26), both sample types show increasing L0 norms, but vulnerable samples maintain their elevated pattern while non-vulnerable samples remain relatively stable.
\end{enumerate}

\subsection{Statistical Analysis}

We conducted statistical tests to verify the significance of the observed differences. A two-sample Kolmogorov-Smirnov test confirms that the distributions of L0 norms between vulnerable and non-vulnerable samples are significantly different (p < 0.001) across all layers.

The effect sizes are particularly pronounced in the middle layers (6-15), where vulnerable samples show 3-5x higher average L0 norms compared to non-vulnerable samples. This suggests that vulnerability detection primarily occurs through intensive computation in these intermediate layers.

To rigorously quantify the differences in internal model representations between vulnerable (TP) and safe (TN) code samples, we analyzed the L2 norms of residual stream activations across all 26 layers of Gemma-2-2b. Following recent advances in mechanistic interpretability and representation engineering, we utilize the L2 norm as a standard metric to track activation magnitudes and standardize latent spaces (e.g., normalizing raw activations such that their expected squared norm is $E[||x||_{2}^{2}]=1$) ~\cite{lieberum2024gemma,ghorbani2019towards,li2023automated}.

The statistical validation includes the following key components:

\begin{itemize}
    \item \textbf{Effect Sizes (Cohen's $d$):} 
    The analysis revealed a distinct phase transition. Early layers (1--3) exhibited strong positive effect sizes (peaking at Layer 2, $d=1.76$), indicating that vulnerable patterns trigger significantly higher activation during initial feature extraction. In contrast, middle layers were largely neutral, while the final output layers inverted to a significant negative effect (Layer 25, $d=-0.74$), suggesting that safe code ultimately results in higher-magnitude semantic representations.

    \item \textbf{Multiple Comparison Corrections:} 
    To control the False Discovery Rate (FDR) across 26 independent layer tests, we applied the Benjamini-Hochberg procedure alongside Bonferroni corrections. The observed differences in the early layers ($p \approx 10^{-26}$) and the final layer ($p \approx 10^{-8}$) remained statistically significant after correction, whereas the activations in the middle reasoning layers (Layers 4--23) were generally indistinguishable between groups in the aggregate analysis.

    \item \textbf{Bootstrap Confidence Intervals:} 
    We computed 95\% confidence intervals (CI) for the mean differences using 1,000 bootstrap resamples. The CI for Layer 2 $[1.06, 1.42]$ strictly excludes zero, demonstrating robust detection of vulnerability features. Conversely, the CI for middle layers (e.g., Layer 12: $[-1.12, 1.29]$) spans zero, reflecting high variance and a lack of consistent directional activation during the contextual integration phase.

    \item \textbf{Per-Vulnerability Breakdown:} 
    Stratifying by CWE revealed mechanism-specific activation fingerprints. Syntactic vulnerabilities, such as Out-of-bounds Write (CWE-787), showed massive early activation ($d=1.99$ at Layer 2) that quickly decayed. However, complex semantic logic errors, such as Double Free (CWE-415), maintained significant activation deep into the network (peaking again at Layer 14, $d=1.10$), suggesting the model utilizes deeper circuits to track variable state and lifecycle information.
\end{itemize}

\subsection{Circuit-Level Findings}
\subsubsection{Critical Attention Heads}

We quantify the importance of attention heads by measuring the divergence in attention concentration (maximum weight and entropy) between vulnerable (TP) and non-vulnerable (TN) samples. 
Table \ref{tab:attention_heads} lists the top-5 heads ranked by importance. 
Notably, all top heads exhibit \textit{negative} importance scores, indicating significantly higher attention activation on non-vulnerable samples. 
This suggests that the model primarily relies on \textbf{Safety Detectors}, i.e., heads that recognize safety patterns (e.g., boundary checks, safe function usage), rather than directly attending to vulnerability features. The absence of these safety patterns triggers the vulnerability classification.

\begin{table*}[h]
  \centering
  \caption{Top-5 Attention Heads Ranked by Importance. Negative scores indicate the head is more active in safe code (Safety Detectors).}
  \label{tab:attention_heads}
  \begin{tabular}{cccl}
    \toprule
    \textbf{Rank} & \textbf{Layer-Head} & \textbf{Importance} & \textbf{Hypothesized Role} \\
    \midrule
    1 & L5-H2 & -0.49 & Safety Syntax Recognition \\
    2 & L2-H2 & -0.42 & Keyword-based Safety Checks \\
    3 & L21-H5 & -0.41 & Semantic Safety Validation \\
    4 & L7-H6 & -0.41 & Boundary Condition Focus \\
    5 & L0-H2 & -0.41 & Initial Token Embedding Focus \\
    \bottomrule
  \end{tabular}
\end{table*}

\begin{figure}[h]
  \centering
\includegraphics[width=\linewidth]{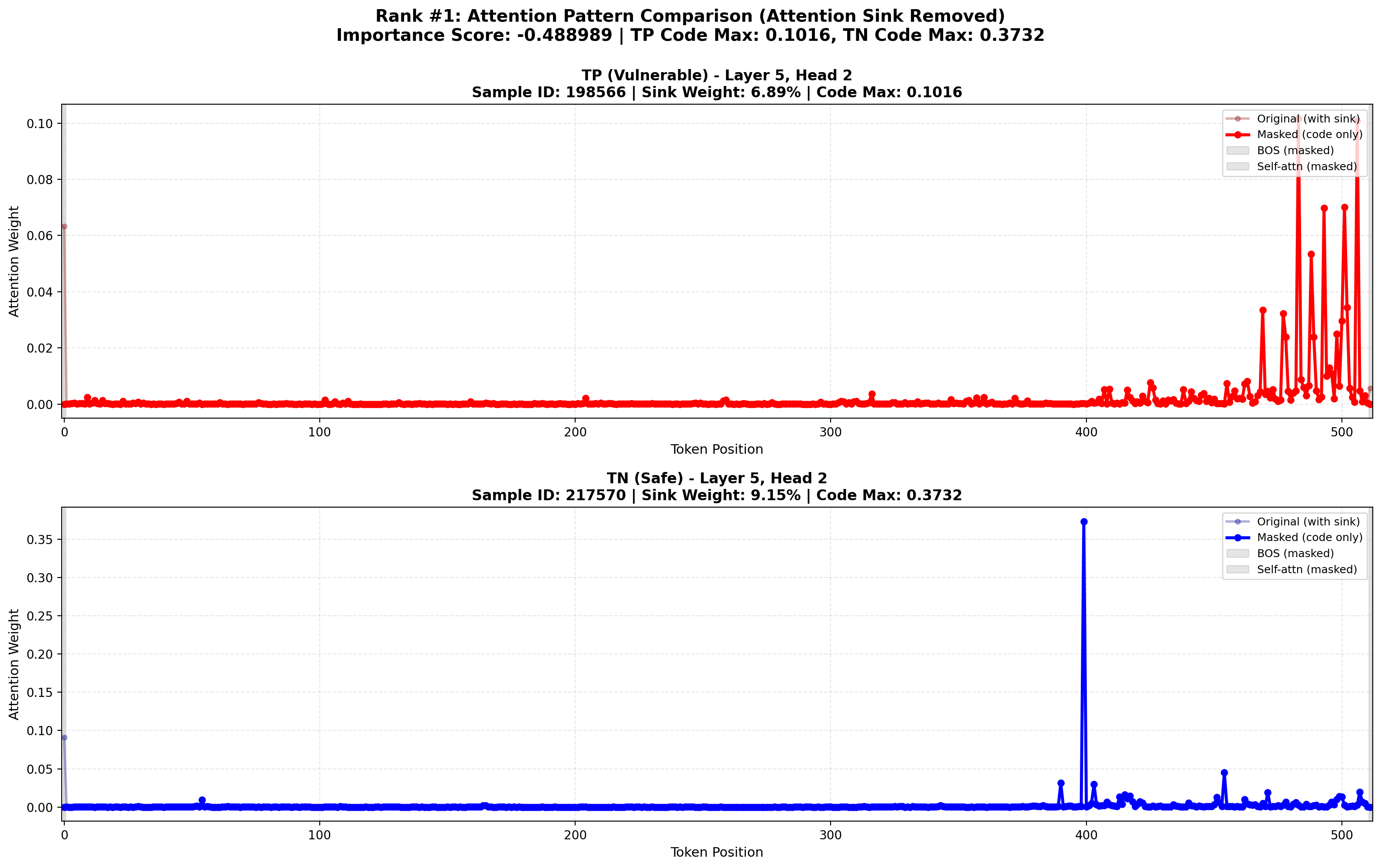}
\caption{Visualization of the attention pattern for the top-ranked head (L5-H2). 
The blue line (Safe/TN) shows sharp attention spikes, indicating the model is focusing on specific safety patterns. 
The red line (Vulnerable/TP) remains flat and diffuse, indicating the absence of such patterns. 
This confirms the head's role as a Safety Detector.}
\label{fig:attn_heatmap}
\end{figure}

As shown in Table \ref{tab:attention_heads}, critical heads are predominantly located in early-to-mid layers (L0, L2, L5, L7), implying that the model's safety judgment is heavily grounded in local syntactic structures rather than deep semantic reasoning.

To validate this hypothesis, we visualized the attention pattern of the top-ranked head (L5-H2), as shown in Figure \ref{fig:attn_heatmap}. 
We observe a distinct contrast: the head exhibits \textbf{sharp attention peaks} on specific tokens in non-vulnerable samples (TN), indicating the successful recognition of safety features. 
In contrast, the attention remains \textbf{diffuse and low-magnitude} in vulnerable samples (TP). 
This "focus-on-safety" behavior provides compelling visual evidence that the head functions as a safety detector, where the absence of strong activation contributes to a vulnerability classification.

\subsubsection{MLP Neuron Specialization}
While attention heads primarily manage information flow and safety validation, we hypothesized that MLPs act as the repository for specific semantic features, including vulnerability patterns. To verify this, we probed the Feed-Forward Networks (FFNs) in intermediate layers (specifically layers 6, 7, 10, and 11) to identify neurons that demonstrate high activation selectivity for vulnerable code.

\begin{figure}[h]
  \centering
  \includegraphics[width=\linewidth]{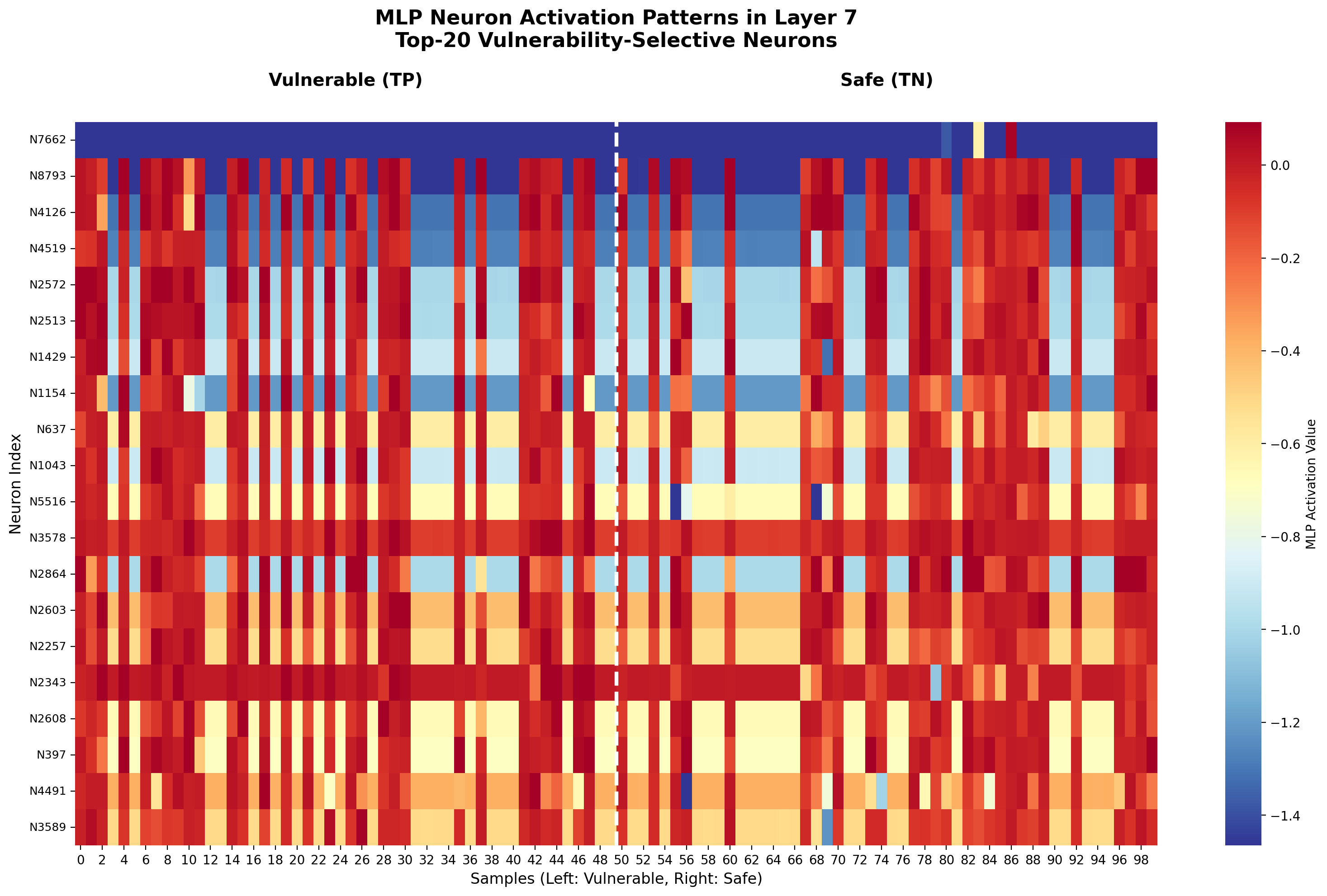}
  \caption{Activation heatmap of the top-20 vulnerability-selective neurons in Layer 7. The x-axis represents individual code samples, split between Vulnerable (left) and Safe (right). The sharp contrast between high activation (red) and suppression (blue) indicates that these neurons specifically encode vulnerability features.}
  \label{fig:neuron_heatmap}
\end{figure}

While the absolute magnitude of selectivity scores (peaking at $\sim$0.25 in Layer 7) may appear modest, it represents a substantial shift in the context of normalized latent spaces.
As visualized in Figure \ref{fig:neuron_heatmap}, this shift is remarkably consistent across samples.
The collective contribution of these specialized neurons provides a strong directional signal to subsequent layers, effectively creating a high-dimensional decision boundary that separates vulnerable from safe representations.

\subsubsection{Computational Circuits}
\begin{figure*}[t]
    \centering
    \includegraphics[width=\linewidth]{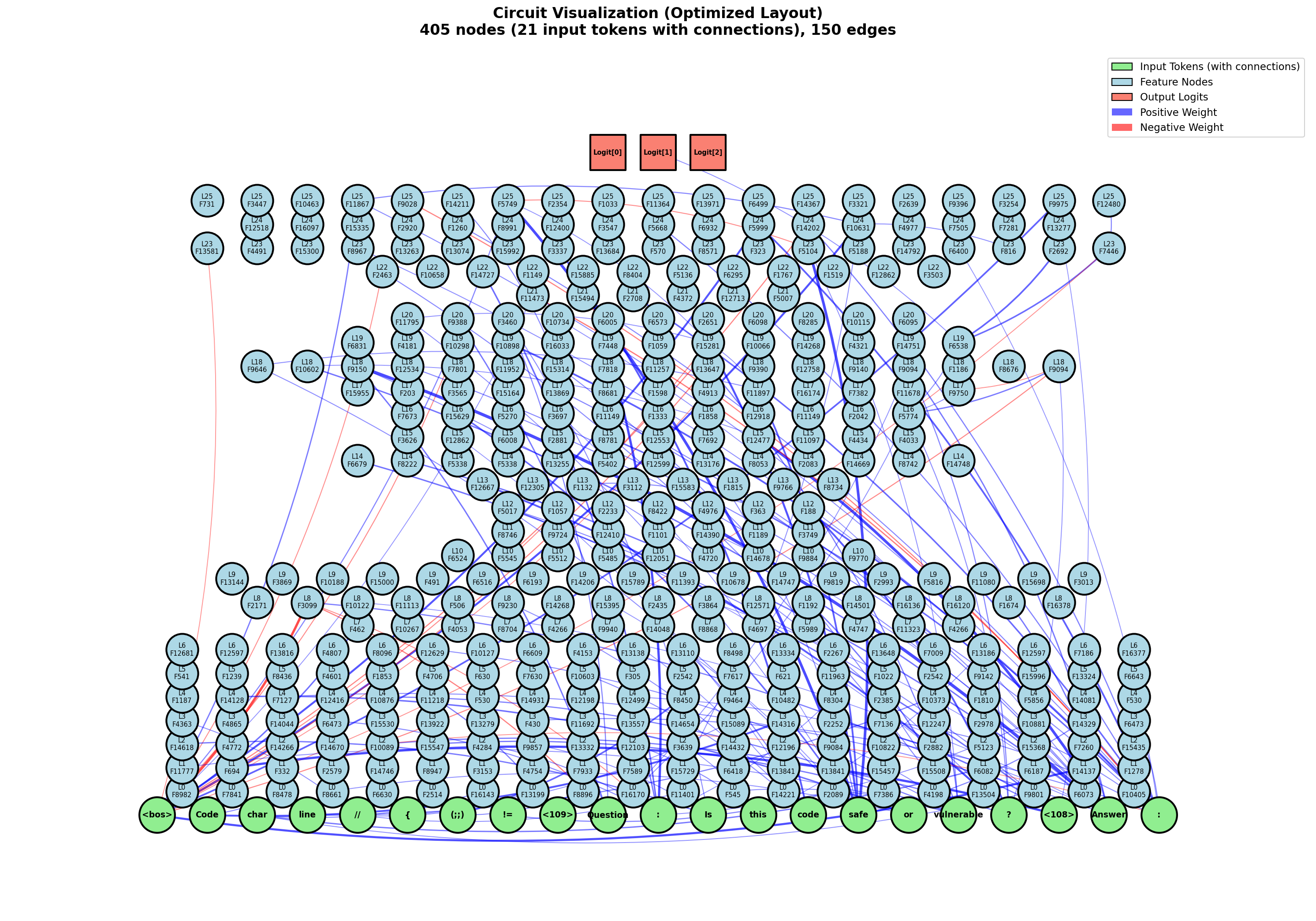} 
    \caption{Visualization of the sparse computational circuit for CWE-787 (Buffer Overflow). The graph traces the information flow from input tokens (bottom, green) through hierarchical feature nodes (middle, blue) to the final output logits (top, red). Edges indicate strong causal attention heads and MLP activations (blue for positive, red for negative weights).}
    \label{fig:circuit_diagram}
\end{figure*}

We applied the Circuit Tracer to map the specific computational pathways activated during vulnerability detection. Figure~\ref{fig:circuit_diagram} illustrates the isolated subgraph for a Buffer Overflow (CWE-787) instance involving an infinite loop construct.

As shown in Figure~\ref{fig:circuit_diagram}, the model does not utilize its full parameter space for detection. Instead, it relies on a highly sparse circuit. Quantitative analysis of the attribution graph reveals that only \textbf{16.1\%} (1,597 out of 9,944) of the probed nodes were active during the inference of the vulnerable sample. 

The information flow exhibits a clear hierarchical structure:
\begin{itemize}
    \item \textbf{Low-Level Syntax Processing (Layers 0--5):} The highest density of active feature nodes is observed in the initial layers (e.g., 414 nodes in Layer 0), where the model processes raw tokens such as \texttt{char\_u} and the infinite loop syntax \texttt{(;;)}.
    \item \textbf{Semantic Aggregation (Layers 6--19):} Activity thins out in the middle layers, suggesting a convergence of syntactic features into higher-level representations.
    \item \textbf{Vulnerability Decision (Layers 20--25):} The final layers show minimal but decisive activity (e.g., only 19 active nodes in Layer 25), where the abstract concept of ``Out-of-bounds Write'' is crystallized before influencing the final output logits.
\end{itemize}

This structure confirms that specific vulnerabilities trigger distinct, localized subgraphs rather than broad, distributed activation patterns across the model.

\subsection{Causal Validation}
To verify the causal role of the identified components, we performed ablation studies on a subset of samples that the model originally classified correctly (Baseline Accuracy = 100\%). We focused on measuring how the removal of specific layers, attention heads, and neurons impacts the model's ability to detect vulnerabilities (True Positive Accuracy).

\subsubsection{Ablation Studies}
\begin{itemize}
    \item \textbf{Layer Criticality:} Ablating Layers 6, 7, and 11 caused a catastrophic drop in vulnerability detection. Specifically, removing Layer 11 resulted in a \textbf{94\% drop} in TP accuracy (from 100\% to 6.02\%), effectively blinding the model to vulnerabilities.
    \item \textbf{Safety Bias:} A distinct ``Safety Default'' behavior was observed. While ablation severely impacted vulnerability detection (TP), the capability to identify safe code (TN) remained relatively robust (e.g., Layer 10 ablation dropped TP to 19.3\% but kept TN at 96.9\%).
    \item \textbf{Neuron Specificity:} We validated the sparsity of the representation by ablating only the \textbf{Top-20 neurons} in Layer 7. This minimal intervention (targeting $<$0.2\% of the layer's neurons) reduced the model's ability to detect vulnerabilities by nearly \textbf{50\%}, confirming that these specific neurons are carriers of the vulnerability signal.
\end{itemize}

\begin{table}[h]
  \caption{Impact of Component Ablation on Detection Accuracy. Note the disproportionate impact on TP (Vulnerability Detection).}
  \label{tab:ablation}
  \centering
  \resizebox{\columnwidth}{!}{%
  \begin{tabular}{lcccc}
    \toprule
    \textbf{Component} & \textbf{Overall} & \textbf{$\Delta$ Overall} & \textbf{TP (Vul)} & \textbf{TN (Safe)} \\
    \midrule
    Baseline (None) & 100.0\% & --- & 100.0\% & 100.0\% \\
    \midrule
    \multicolumn{5}{l}{\textit{Layer-wise Mean Ablation}} \\
    Layer 6 & 76.4\% & -23.6\% & 27.7\% & 94.6\% \\
    Layer 7 & 51.2\% & -48.8\% & 19.3\% & 63.1\% \\
    Layer 10 & 75.7\% & -24.3\% & 19.3\% & 96.9\% \\
    Layer 11 & 48.9\% & -51.1\% & \textbf{6.0\%} & 64.9\% \\
    \midrule
    \multicolumn{5}{l}{\textit{Fine-grained Ablation}} \\
    L7 (Top-20 Neurons) & 82.3\% & -17.7\% & 50.6\% & 94.1\% \\
    L0 H2 (Attention) & 85.9\% & -14.1\% & 61.5\% & 95.1\% \\
    \bottomrule
  \end{tabular}%
  }
\end{table}

\subsubsection{Activation Patching}

We performed causal activation patching (also known as activation steering) to localize the specific layers responsible for vulnerability detection. Specifically, we computed a mean activation vector from non-vulnerable (safe) samples and injected it into vulnerable samples to observe if the model's prediction would flip from "vulnerable" to "safe".

\begin{figure}[h]
    \centering
    \includegraphics[width=0.9\linewidth]{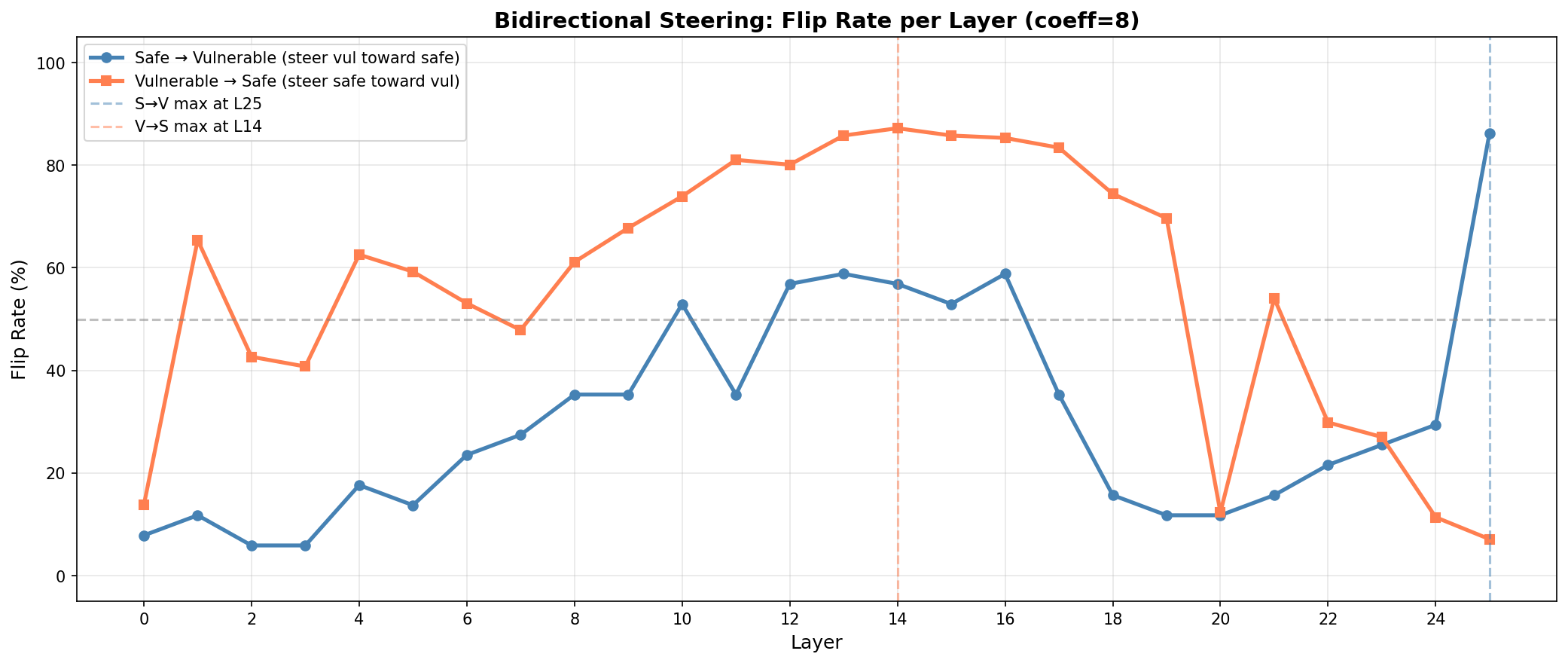} 
    \caption{Layer-wise flip rates using bidirectional causal patching. Injecting safe vectors into vulnerable samples (blue line) peaks at Layer 14, indicating this is the critical decision-making layer.}
    \label{fig:patching}
\end{figure}

Our experiments reveal three key findings:
\begin{itemize}
    \item \textbf{Critical Decision Layer:} Patching early layers (0--9) has minimal impact on the model's output (flip rate $<20\%$), indicating that these layers primarily process low-level syntax. However, the flip rate rises significantly in the middle layers, peaking at \textbf{Layer 14} (reaching over 45\% flip rate with standard coefficient). This suggests that high-level vulnerability semantics are aggregated and formed around Layer 14.
    
    \item \textbf{Asymmetry in Representation:} We observed an asymmetry between the two patching directions. While injecting safe vectors into vulnerable samples (Safe $\to$ Vulnerable) effectively flips predictions at Layer 14, the reverse direction (Vulnerable $\to$ Safe) requires a significantly higher steering coefficient to achieve similar flip rates. This implies that the model has a robust internal prior for "safety," while "vulnerability" features are more specific and sparse.
    
    \item \textbf{Robustness Analysis:} By increasing the steering coefficient (e.g., coeff=8), we confirmed that the decision mechanism is consistently localized to the middle layers (Layers 12--16), validating Layer 14 as the semantic hub for security analysis.
\end{itemize}

\subsection{Vulnerability Type Analysis}

We categorized samples according to CWE labels from the PrimeVul dataset to analyze whether different vulnerability types activate distinct circuits.

\begin{figure}
  \includegraphics[width=\columnwidth]{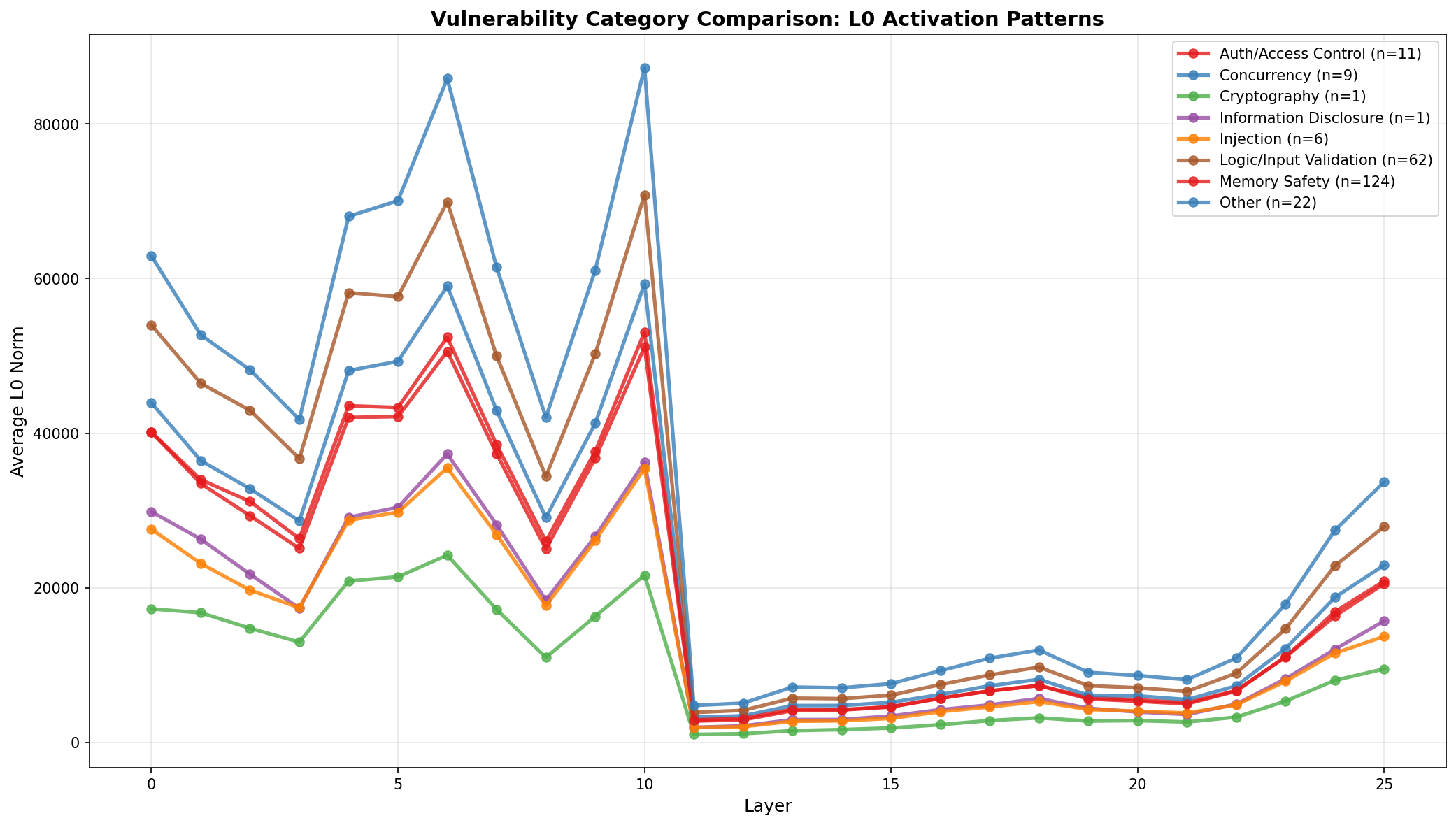}
  \caption{L0 activation patterns across vulnerability categories. Different 
  vulnerability types show distinct activation signatures, particularly in 
  layers 5-10. Memory Safety (n=124) and Concurrency (n=9) vulnerabilities 
  trigger significantly higher activations compared to other categories.}
  \label{fig:vuln_types}
\end{figure}

Different vulnerability categories activate distinct computational circuits with dramatically different feature requirements 
(Figure~\ref{fig:vuln_types}). Concurrency vulnerabilities trigger 70\% 
higher L0 activation than memory safety issues, suggesting fundamental 
differences in detection complexity.

\section{Discussion}

\subsection{Mechanistic Understanding of Vulnerability Detection}

Our Circuit Tracer analysis reveals that vulnerability detection in LLMs is not a holistic, distributed process but rather involves specific, identifiable computational circuits. The concentration of critical activations in layers 6-7 and 10-11 suggests a hierarchical processing pipeline.


\subsubsection{Safety Detection as Primary Mechanism}

A surprising finding from our attention head analysis is that the model primarily relies on \textit{safety detectors} rather than vulnerability detectors. The top-ranked attention heads (L5-H2, L2-H2, L7-H6) all exhibit negative importance scores, indicating they activate strongly for safe code patterns rather than vulnerable ones. This suggests the model learns to recognize secure coding practices (boundary checks, proper input validation, safe API usage) and classifies code as vulnerable when these safety patterns are absent.

This ``detection by absence'' mechanism has important implications: the model's vulnerability detection is fundamentally conservative, defaulting to ``vulnerable'' when it cannot identify familiar safety patterns. This explains the observed ``Safety Default'' behavior in our ablation studies, where removing critical layers disproportionately impacts vulnerability detection (TP drops to 6\%) while safe code recognition remains relatively robust (TN at 64.9\%).

\subsubsection{Early-Middle Layers (6-7): Feature Encoding}

Our MLP neuron analysis reveals that Layer 7 contains highly specialized neurons for vulnerability-related features. The top-20 neurons identified by our Vulnerability Selectivity Score show consistent activation patterns across vulnerable samples while remaining suppressed for safe code (Figure~\ref{fig:neuron_heatmap}). Ablating just these 20 neurons (less than 0.2\% of the layer) reduces vulnerability detection by nearly 50\%, confirming their critical role as feature encoders.

The attention heads in these layers (particularly L5-H2 and L7-H6) focus on local syntactic structures, suggesting they process code tokens to identify safety-relevant patterns such as bounds checking and null pointer validation.

\subsubsection{Middle-Late Layers (10-11): Decision Crystallization}

Layers 10-11 show the most catastrophic impact when ablated: Layer 11 ablation drops TP accuracy from 100\% to just 6\%. This suggests these layers perform the critical computation that transforms distributed feature representations into the final vulnerability/safe decision. The sparse circuit analysis (Figure~\ref{fig:circuit_diagram}) confirms this interpretation, showing that activity converges dramatically in the final layers (only 19 active nodes in Layer 25) where the abstract vulnerability concept is crystallized.

\paragraph{Specific Computation: Vulnerability Assertion.}
We further hypothesize that Layers 10-11 function as a specific \textit{``Vulnerability Assertor.''} This is evidenced by the stark asymmetry in our ablation results (Table~\ref{tab:ablation}): removing Layer 11 obliterates the model's ability to detect vulnerabilities (TP drops to 6.0\%) while partially preserving its ability to recognize safe code (TN remains at 64.9\%).
This indicates that the default state of the information flow is neutral or safe, and Layer 11 is critically responsible for actively asserting the presence of a flaw by aggregating the sparse feature signals (identified in Layer 7) into a cohesive semantic judgment. Without this integration step, the isolated risk features from earlier layers fail to trigger a detection outcome.


\subsection{Circuit Sparsity and Modularity}

Our circuit tracing reveals that vulnerability detection uses only \textbf{16.1\%} of the model's probed nodes, demonstrating remarkable sparsity. This finding has two important implications:

First, it suggests that vulnerability detection is a \textit{localized} rather than distributed computation. The model does not engage its full capacity but instead routes information through specific, identifiable pathways. This modularity makes the circuits amenable to targeted intervention and interpretation.

Second, the hierarchical structure, from dense early-layer processing (414 nodes in Layer 0) to sparse late-layer decision-making (19 nodes in Layer 25), mirrors the abstraction hierarchy in human code analysis: lexical parsing → pattern recognition → semantic judgment.

\subsection{Vulnerability-Specific Circuits}

Different vulnerability categories activate distinct computational circuits with dramatically different feature requirements (Figure~\ref{fig:vuln_types}). Concurrency vulnerabilities trigger 70\% higher L0 activation than memory safety issues, suggesting that detecting race conditions and synchronization bugs requires more intensive computation than identifying buffer overflows.

Memory safety vulnerabilities often manifest as local syntactic patterns (e.g., a \texttt{strcpy} call without a length check). The relatively low $L_0$ activation suggests the model resolves these using \textit{sparse, localized circuits} focused on pattern matching in early layers.

In contrast, concurrency bugs involve complex interactions between threads and shared resources, which are non-local and context-dependent. The elevated activation levels indicate that the model must engage \textit{denser, distributed circuits} to track state changes and logic flow, effectively allocating more computational resources to reason about the flaw.

This confirms that the ``Vulnerability Detection Circuit'' is not a static subgraph but a \textit{dynamic mechanism} that scales its computational resource allocation according to the semantic complexity of the code defect.

Synthesizing our findings from the attention head analysis and activation patterns, we propose that the model employs a \textit{``Core-Periphery''} architecture for detection:
\begin{itemize}
    \item \textbf{The Shared Core:} The identified ``Safety Detectors'' (specifically heads L5-H2 and L7-H6) function as a universal gatekeeper. These components are active across varying vulnerability types, tasked with identifying fundamental safe coding patterns (e.g., boundary checks) regardless of the specific threat model.
    \item \textbf{Vulnerability-Specific Extensions:} When the shared core fails to validate safety, the model recruits specialized sub-circuits. As evidenced by the distinct activation costs (Figure~\ref{fig:vuln_types}), simple CWEs rely on shallow pattern matching, whereas complex logic errors (e.g., Concurrency) trigger deeper, denser computation pathways.
\end{itemize}

This disentanglement of general safety checks from specific vulnerability logic supports the development of modular, lightweight detectors.Since the vulnerability detection circuit is sparse , it is theoretically possible to prune the model into specialized sub-networks. Instead of deploying a monolithic LLM, security systems could utilize a cascade of smaller, extracted circuits—running the cheap shared core first, and only activating the expensive specific extensions when potential logic flaws are suspected.


\subsection{Implications for Security Applications}

Our mechanistic insights have several practical implications:

\subsubsection{Circuit-Level Explanations}

Unlike black-box vulnerability detectors that provide only binary outputs, our analysis enables circuit-level explanations. For a given prediction, we can now identify which attention heads activated (or failed to activate), which MLP neurons contributed to the decision, and trace the information flow through the model. For example, if L5-H2 (Safety Syntax Recognition) shows weak activation, we can explain that the model did not detect expected safety patterns in the code.

\subsubsection{Targeted Model Improvement}

Understanding circuit structure enables targeted interventions. Our ablation results identify Layer 11 as the critical bottleneck, and any improvement effort should prioritize this layer. Similarly, the 20 vulnerability-selective neurons in Layer 7 represent high-value targets for fine-tuning or augmentation. Rather than retraining the entire model, practitioners could focus on strengthening these specific circuits.

\subsubsection{Failure Mode Analysis}

The ``Safety Default'' behavior revealed by our ablations suggests a specific failure mode: the model may produce false positives when encountering unfamiliar but safe coding patterns that do not trigger its learned safety detectors. This insight can guide the collection of training data to include diverse safe coding styles.

Our discovery that the model relies on \textit{Safety Detectors} (e.g., L5-H2) implies a specific vulnerability to evasion. Since the model classifies code as vulnerable primarily when safety patterns are \textit{absent}, attackers could theoretically craft \textit{``Safety Mimicry''} attacks. By injecting inert but syntactically correct safety checks (e.g., irrelevant boundary validations) into malicious code, an adversary could artificially activate these safety heads, forcing a False Negative. This highlights the inherent fragility of the model's ``detection by absence'' mechanism compared to positive vulnerability recognition.


\subsection{Comparison with Human Reasoning}

The hierarchical processing pipeline we identified (syntactic pattern detection followed by semantic aggregation and decision crystallization) parallels how human security experts analyze code. Experts first scan for suspicious patterns (dangerous functions, missing checks), then reason about data flow and control flow, and finally make a judgment. The model appears to have learned a similar multi-stage process, though it relies on ``safety detection'' rather than direct vulnerability recognition.

Interestingly, this safety-first approach may reflect the training data distribution: secure code is far more common than vulnerable code, so the model learns robust representations of safety patterns while treating their absence as the vulnerability signal.

Our finding that the model primarily utilizes \textit{Safety Detectors} (e.g., L5-H2) suggests a distinct alignment with specific human analysis strategies. Unlike automated scanners that typically employ ``Bug Hunting'' (searching for negative signatures of known exploits), the LLM appears to mimic a ``Compliance Verification'' strategy often used by human auditors: actively scanning for the presence of necessary safeguards (e.g., input sanitization) and flagging code where these expected patterns are absent.

While humans categorize vulnerabilities by consequence (e.g., CWE labels), our circuit analysis suggests the model may discover novel patterns based on computational complexity. The significant divergence in activation magnitude between Memory Safety and Concurrency vulnerabilities implies the model constructs a latent taxonomy based on ``reasoning depth''—distinguishing between local syntactic flaws and global state-management errors—which offers a complementary perspective to rigid human taxonomies.


\subsection{Limitations and Future Work}

Our analysis has several limitations in scope. All experiments use Gemma-2-2b, and vulnerability detection circuits may differ substantially in other architectures such as CodeBERT, StarCoder, or GPT-based models. The dataset of 472 samples, while sufficient for initial circuit discovery, limits statistical power for fine-grained per-CWE analysis. Additionally, our analysis is restricted to C/C++ code and covers only 9 CWE categories; circuits for other programming languages and vulnerability types may exhibit different characteristics.

Several methodological extensions would strengthen these findings. Sparse autoencoders could decompose the polysemantic neurons we identified in Layers 7 and 11 into monosemantic features, providing cleaner interpretations of what each neuron encodes. Cross-model analysis comparing circuits across different LLMs would reveal whether the safety-detector mechanism is universal or architecture-specific. Activation patching experiments could systematically identify at which layer the vulnerability representation forms, complementing our ablation results. Finally, synthetic probing using minimal code examples that maximally activate specific safety detectors would provide more direct evidence for the features these circuits recognize.

\subsection{Broader Impact}

Understanding the internal mechanisms of vulnerability detection systems carries both benefits and risks. On the positive side, interpretable circuits enable more trustworthy security tools by allowing practitioners to verify why a model flagged specific code. This transparency supports debugging and targeted improvement of detection systems, and facilitates regulatory compliance in domains requiring explainable AI.

However, exposing model internals also creates potential for misuse. Adversaries could craft evasion techniques that specifically target identified circuit weaknesses, or exploit knowledge of the model's blind spots to bypass detection. The ``safety detector'' mechanism we identified suggests a concrete attack surface: code that avoids triggering learned safety patterns while remaining vulnerable.

We believe the benefits of transparency outweigh these risks for security-critical applications, where understanding model failures is essential for responsible deployment. That said, production systems should incorporate adversarial testing informed by circuit-level findings and continuous monitoring for evasion attempts.

\section{Conclusion}

This study provides the first mechanistic analysis of computational circuits underlying LLM-based vulnerability detection. Through Circuit Tracer instrumentation of Gemma-2-2b across 472 C/C++ code samples and 26 transformer layers, we have uncovered the specific neural components responsible for identifying vulnerable code patterns.

Our central finding is that the model relies on \textit{safety detectors} rather than vulnerability detectors. The top-ranked attention heads (L5-H2, L2-H2, L7-H6) all exhibit negative importance scores, activating strongly for safe code patterns such as boundary checks and proper input validation. When these safety patterns are absent, the model classifies code as vulnerable. This ``detection by absence'' mechanism explains the asymmetric impact of ablations: removing Layer 11 drops vulnerability detection from 100\% to just 6\%, while safe code recognition remains at 64.9\%.

We identified a hierarchical two-stage processing pipeline. Early-middle layers (6-7) encode vulnerability-relevant features through specialized MLP neurons; ablating just 20 neurons in Layer 7 (less than 0.2\% of the layer) reduces vulnerability detection by 50\%. Middle-late layers (10-11) perform decision crystallization, transforming distributed features into the final classification. Circuit tracing reveals remarkable sparsity: only 16.1\% of probed nodes are active during vulnerability detection, with activity converging from 414 nodes in Layer 0 to just 19 nodes in Layer 25.

Different vulnerability categories activate distinct circuits with varying computational requirements. Concurrency vulnerabilities trigger 70\% higher L0 activation than memory safety issues, suggesting that detecting race conditions requires more intensive computation than identifying buffer overflows.

These findings enable circuit-level explanations for security predictions, targeted model improvements focused on critical components like Layer 11 and the vulnerability-selective neurons in Layer 7, and better understanding of failure modes such as false positives on unfamiliar safe coding patterns. Future work should extend this analysis to other model architectures and employ sparse autoencoders to decompose the polysemantic neurons we identified.



\section*{Data Availability Statement}
The dataset, instrumentation code, and analysis scripts necessary to reproduce the findings in this paper are available in a public repository at \url{https://anonymous.4open.science/r/LLMvul-02E6/}. To facilitate open science prior to final publication, this repository contains the complete experimental artifact. The full artifact, including all fine-tuned model weights and extended datasets, will be formally archived via a permanent DOI (Zenodo) and Hugging Face upon formal publication.

\begin{acks}
This work is supported by the Wallenberg AI, Autonomous Systems and Software Program (WASP) funded by the Knut and Alice Wallenberg Foundation. The computations and data handling in this study were enabled by high-performance computing resources provided by LUNARC, The Centre for Scientific and Technical Computing at Lund University.
\end{acks}

\bibliographystyle{ACM-Reference-Format}
\bibliography{llm_vulnerability}


\end{document}